\documentclass[aps,prb,twocolumn,showpacs,preprintnumbers,amsmath,amssymb,superscriptaddress]{revtex4}

\usepackage{graphicx}
\usepackage{dcolumn}
\usepackage{bm}
\usepackage{figsize}

\begin{document}


\title{Large-scale impurity potential in the quantum Hall effect for the HgTe quantum well with inverted band structure }

\author{S.~V.~Gudina}
\affiliation{M.N. Miheev Institute of Metal Physics of Ural Branch of Russian Academy of Sciences, 18 S. Kovalevskaya Str., Ekaterinburg 620990, Russia}
\author{Yu.~G.~Arapov}
\affiliation{M.N. Miheev Institute of Metal Physics of Ural Branch of Russian Academy of Sciences, 18 S. Kovalevskaya Str., Ekaterinburg 620990, Russia}
\author{V.~N.~Neverov}
\email[]{neverov@imp.uran.ru}
\affiliation{M.N. Miheev Institute of Metal Physics of Ural Branch of Russian Academy of Sciences, 18 S. Kovalevskaya Str., Ekaterinburg 620990, Russia}
\author{E.~G.~Novik}
\affiliation{ Physical institute, University of W\"{u}rzburg, D-97074 W\"{u}rzburg, Germany}
\author{S.~M.~Podgornykh}
\affiliation{M.N. Miheev Institute of Metal Physics of Ural Branch of Russian Academy of Sciences, 18 S. Kovalevskaya Str., Ekaterinburg 620990, Russia}
\affiliation{Ural Federal University, 19 Mira Str., Ekaterinburg 620002, Russia}
\author{M.~R.~Popov}
\affiliation{M.N. Miheev Institute of Metal Physics of Ural Branch of Russian Academy of Sciences, 18 S. Kovalevskaya Str., Ekaterinburg 620990, Russia}
\author{E.~V.~Ilchenko}
\affiliation{M.N. Miheev Institute of Metal Physics of Ural Branch of Russian Academy of Sciences, 18 S. Kovalevskaya Str., Ekaterinburg 620990, Russia}
\author{N.~G.~Shelushinina}
\affiliation{M.N. Miheev Institute of Metal Physics of Ural Branch of Russian Academy of Sciences, 18 S. Kovalevskaya Str., Ekaterinburg 620990, Russia}
\author{M. V. Yakunin}
\affiliation{M.N. Miheev Institute of Metal Physics of Ural Branch of Russian Academy of Sciences, 18 S. Kovalevskaya Str., Ekaterinburg 620990, Russia}
\affiliation{Ural Federal University, 19 Mira Str., Ekaterinburg 620002, Russia}
\author{N. N. Mikhailov}
\affiliation{Rzhanov Institute of Semiconductor Physics Sibirean Branch of Russian Academy of Sciences, 13 Lavrentyev Ave., Novosibirsk 630090, Russia}
\author{S. A. Dvoretsky}
\affiliation{Rzhanov Institute of Semiconductor Physics Sibirean Branch of Russian Academy of Sciences, 13 Lavrentyev Ave., Novosibirsk 630090, Russia}

\date{\today}

\begin{abstract}
We report on the longitudinal and Hall resistivities of a HgTe quantum well with inverted energy spectrum ($d_{QW} = 20.3$ nm) measured in the quantum Hall (QH) regime at magnetic fields up to 9 T and temperatures 2--50 K. The temperature dependence of the QH plateau-plateau transition (PPT) widths and of variable range hopping (VRH) conduction on the Hall plateaus are analyzed. The data are presented in a genuine scale form both for PPT regions and for VRH regime. Estimations for the degree of the carrier localization length divergence reveal a decisive role of the long-range random potential (the potential of remote ionized  impurities) in the localization- delocalization processes in the QH regime for the system under study.
\end{abstract}

\pacs{73.21.Fg, 73.43.-f, 73.43.Qt, 73.43.Nq}
\maketitle

\section{\label{sec:1}Introduction}

A remarkable property of the HgTe-based quantum well (QW) structures is that transitions between band insulator (BI),  topological insulator (TI) and semimetal (SM) phases may be achieved by tuning the quantum-well thickness $d_{QW}$ \cite{Konig,Bernevig,Konig-2007Science,Krishtopenko,Kvon-JETPLett2008,Kvon-PRB2011} (see, for example, Fig.1 in Ref.~\onlinecite{Krishtopenko}). The ordinary 2D insulator state is realized at small well widths (up to a critical thickness $d_C \approx 6.3$nm) \cite{Konig}, while 2D TI exists at larger well widths (up to $d_{QW} \approx 14$nm) \cite{Bernevig,Konig-2007Science,Krishtopenko}. For the width $d_{QW} \approx (18-20)$ nm and wider quantum wells SM 2D state with overlapped conduction and valence 2D bands is realized \cite{Krishtopenko,Kvon-JETPLett2008,Kvon-PRB2011}.
A clear model for the physics of the relevant subbands of HgTe/CdTe QW based on the bulk HgTe and CdTe band structure is presented in Ref.~\onlinecite{Konig}.

It is well-known that both HgTe and CdTe bulk materials have the zinc-blende lattice structure where important bands are close to the $\Gamma$-point in the Brillouin zone, and they are the s-type band ($\Gamma_6$) and the $p$-type band, which is split to a $J = 3/2$ -band ($\Gamma_8$) and a $J = 1/2$ -band ($\Gamma_7$) by spin-orbit coupling.

CdTe has a "normal" band order with $\Gamma_6$ conduction band and ($\Gamma_8$; $\Gamma_7$) valence bands. The highest valence band $\Gamma_8$ is separated from the conduction band by a large energy gap $\varepsilon_g = 1.6$eV ($\varepsilon_g \equiv E(\Gamma_6)-E(\Gamma_8)$). 

In a bulk HgTe due to relativistic effects \cite{Tsidilkovski} the $\Gamma_8$ band, which "normally" forms the valence band, is now above the $\Gamma_6$ band that indicates a negative energy gap $\varepsilon_g=-300$meV. The light-hole bulk subband of the $\Gamma_8$ band becomes the conduction band and the heavy-hole bulk subband becomes the first valence band. Based on this unusual sequence of the $\Gamma_6$ and $\Gamma_8$ states, such a band structure is called "inverted". 

When Cd(Hg)Te/HgTe/Cd(Hg)Te structures with  HgTe quantum well (QW) are grown, for a thin QW layer the quantum confinement gives rise to the "normal" sequence of subbands, similar to CdTe, i.e., the bands with primarily $\Gamma_6$ symmetry are the conduction subbands and the $\Gamma_8$ bands contribute to the valence subbands (BI phase).

As the QW thickness is increased, the material looks more like HgTe and for wide QW layers the band structure tends to be "inverted". The inverted regime is achieved when QW width, $d_{QW}$, exceeds a critical value $d_c \cong 6.3$nm.
At $d_{QW} = d_c$ the conduction and valence bands touch each other, which leads to a single-valley gapless 2D Dirac-fermion system \cite{Buttner,KozlovJETPL2015} where the quantum Hall effect (QHE) can be observed up to nitrogen temperatures \cite{KozlovAPL2014}.

At the critical thickness $d_c$ a topological phase transition from a 2D BI with "normal" band ordering to a 2D TI with an "inverted" one occurs \cite{Krishtopenko}. Recent years demonstrate astounding growth in research on topological insulators, the materials that have a bulk band gap like an ordinary insulator but support conducting states on their edge , so-called quantum spin Hall (QSH)  states \cite{QiRevModPhys2011-11,BernevigPRL2006-12,GusevSSC2015-13}. 

The first 2D TIs discovered were based on HgTe/Cd(Hg)Te quantum wells \cite{Bernevig,Konig-2007Science}. The origin of the 2D TI phase in HgTe/Cd(Hg)Te systems is caused by a peculiar size quantization  for HgTe QWs  with the inverted band structure \cite{Konig}.

The gap between the ground-state heavy hole  subband (H1) and the next adjacent subband exists for QW narrower than 18 nm (for wider wells the QW is in a SM state). While the gap is open, a HgTe-based QW should be a 2D TI having edge states in the gap between subbands . Since the first works \cite{Bernevig,Konig-2007Science} and up to now the 2D-TI  is the most studied and trendy domain for the HgTe based heterostructures (see, for  example, \cite{Krishtopenko,GusevPRB2013-14,GusevPRB2013-15,OlshanetskyPRL2015-16,DantscherCond-mat16-17} and references therein).

In wide HgTe/Cd(Hg)Te QWs with an inverted energy band structure ($d_{QW} \gtrsim 18$nm) a novel 2D electron system has been shown to exist: a 2D SM \cite{Kvon-JETPLett2008,Kvon-PRB2011,ZholudevPRB2012-18}. The existence of 2D SM in this system is due to the overlap by about a few meV of the conduction band minimum at the center of the
Brillouin zone with the side  maxima of the valence band. Calculation of the energy band structure \cite{Kvon-PRB2011} shows that a key reason for the overlap in wide QWs is the strain due to the lattice mismatch between HgTe and CdTe.

When the Fermi level crosses both the valence and conduction bands, a number of interesting transport properties caused by the simultaneous presence of 2D electrons and holes have been observed  in HgTe QWs.

In SM domain (since the first works of Kvon et al. \cite{Kvon-JETPLett2008,Kvon-PRB2011}) the emphasis is made on a classical magnetoresistance, Shubnikov - de Haas oscillation (SHO) pattern and QHE for two \cite{Kvon-JETPLett2008,Kvon-PRB2011,GusevPRL2010-19} or even three \cite{Minkov-2013-20,YakPRB-2016-21} types of carriers in a single \cite{Kvon-JETPLett2008,Kvon-PRB2011,GusevPRL2010-19,Minkov-2013-20} or in a double \cite{YakPRB-2016-21} QW, in weakly doped  structures \cite{Kvon-JETPLett2008} or at different density ratios of the two dimensional electrons and holes in the structures with an applied gate voltage $V_g$ \cite{Kvon-PRB2011,GusevPRL2010-19,Minkov-2013-20,YakPRB-2016-21}.

We present a study of quantum magnetotransport in a 20.3-nm-wide HgTe QW grown on the (013) GaAs substrate, symmetrically modulation doped with In at both sides of QW. 
Formally, we are in a SM phase but doping with In ensures the Fermi level position in the conduction band above the lateral maximum of the valence band. Because of this we observe an ordinary picture of QHE for one type of carriers (electrons) that allows us to investigate more subtle effects of localization - delocalization in the QHE regime. It is an identification of scaling conditions both for the quantum phase plateau-plateau transition and for the variable-range hopping conductivity on the localized states at the Hall plateaus.

QHE plateau-plateau transition, as well as the plateau-insulator transition, in high quality HgTe QW with an inverted band structure were first studied and analyzed within scaling concepts at $T= (0.3 - 3.0)$K \cite{OlshanetskyJETPL2006-22} where it was concluded that the applicability of scaling models to this system is problematic.

In our previous work \cite{ArapovSemic2015-23}, we have presented the data on the temperature dependence of  the PPT width, $\nu_0(T)$, for a HgTe quantum well with inverted energy spectrum ($d_{QW} = 20.3$nm). The actual scaling behavior $\nu_0(T)\sim T^\kappa$ is observed for the $1 \to 2$ PPT in a wide temperature range $T =(2.9-30)$K. 

Recently, using the scaling approach for the $1 \to 2$ PPT in  non-inverted  HgTe QW ($d_{QW} = 5.9$nm) , Khouri et al. \cite{KhouriPRB-2016-24} have found an excellent agreement with the universal scaling theory: the scaling coefficient $\kappa =(0.45 \pm 0.04)$ at $T = (0.3-60)$K.

As for the variable-range hopping conductivity in the minima of  $\sigma_{xx}$, associated with the Hall plateau regions, it is a widely used method for a detection of the localization length divergence in QHE regime at a number of 2D systems (see, for example, Refs [10-18] in Ref.~\onlinecite{ArapovJLTP2016-25}). But for the HgTe based 2D system, this method was first used by us \cite{ArapovJLTP2016-25,GudinaPSSC2016-26}: an analysis of the VRH conductivity in the regions of the first and second quantum Hall plateaus provided an opportunity to determine the value and the magnetic-field dependence of the localization length  in the HgCdTe/HgTe/HgCdTe heterostructure with a wide HgTe quantum well. 

The objectives of this work are:
\begin{itemize}
\item to generalize the data obtained by us in Ref.~\onlinecite{ArapovSemic2015-23} and in Ref.~\onlinecite{ArapovJLTP2016-25,GudinaPSSC2016-26} by presenting them in a genuine scale form both for PPT regions and for VRH regime;
\item to dovetail our results on PPT and on VRH with a general picture of studies the localization effects in QHE  for systems both with short-range and large-scale random impurity potential;
\item  in particular, to compare our results with the data for modulation-doped GaAs/GaAlAs systems.
\end{itemize}

The analysis has led us to the conclusion that, similarly to modulation-doped GaAs, for PPT in our HgTe QW we are in the intermediate between quantum tunneling and classical percolation region of localization length divergence. On the other hand, the VRH regime in our system is realized by hopping between localized states in the tails of Landau levels, which is within the scope of the laws for classical percolation outside the region of quantum tunneling.

All this indicates the decisive role of the long-range random potential (the potential of remote ionized In impurities) for scattering and localization of carriers in the system under study.

\section{\label{sec:2}Characteristics of the sample}

The sample is a 20.3-nm-wide HgTe quantum well between Hg$_{0.35}$Cd$_{0.65}$Te barriers grown on the (013) GaAs substrate, symmetrically modulation doped with In at both sides at distances of about 10 nm spacers. The electron gas density is $n_s = 1.5\times 10^{15}$m$^{-2}$ with a mobility of 22 m$^2$/Vs. The sample is in the shape of a Hall bar with Ohmic contacts.

\subsection{\label{sec:21}Band structure of the HgTe quantum well}

\begin{figure}[b]
\includegraphics[width=90mm]{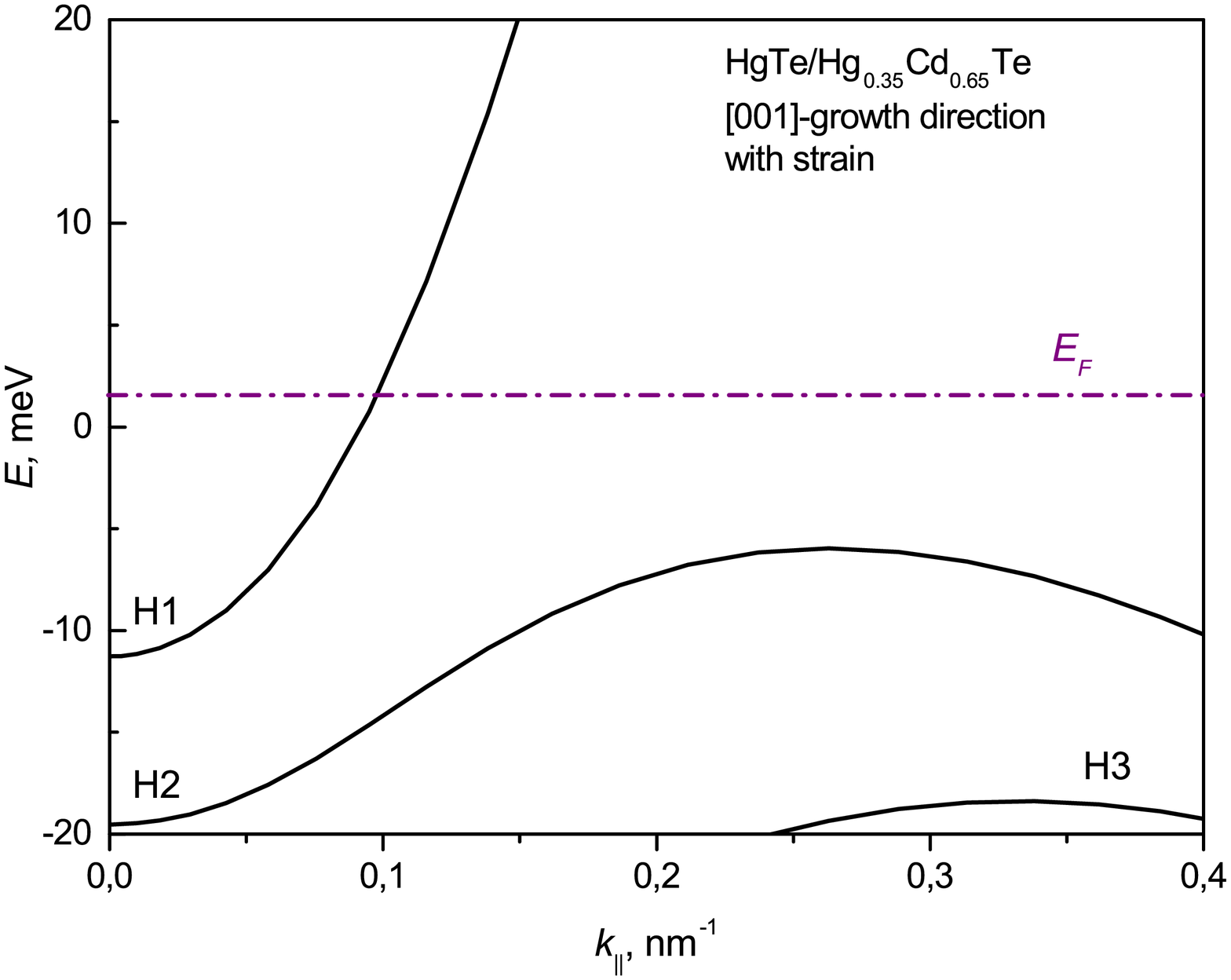}
\caption{\label{fig1:BandStrucHgTe} The band structure of a HgTe quantum well with $d_{QW}=20$nm calculated for (001) orientation. Dot-dashed horizontal line shows $E_F$ position for $n = 1.5\times 10^{15}$m$^{-2}$.}
\end{figure}

The subband energy dispersion for a fully strained 20 nm-wide HgTe QW in Hg$_{0.35}$Cd$_{0.65}$Te/HgTe/Hg$_{0.35}$Cd$_{0.65}$Te nanostructure is shown in Fig.~\ref{fig1:BandStrucHgTe} for the (001) orientation. We suppose that differences between the calculated (001) and experimental (013) orientations, although introduce some quantitative corrections, would not considerably influence the results of present study. Calculations are performed in an envelope function approach within the framework of 8-band $\mathbf k\times \mathbf p$ theory self-consistently with the Poisson equation for the charge distribution \cite{NovikPRB-2005-27}.
In the inverted regime of HgTe QW the first size-quantized heavy hole subband H1 becomes the lowest conduction band as the theory \cite{DyakonovJETP-1982-28,GerchikovPSSB-1990-29} predicts for it an electron-like effective mass. The highest valence band is now the second size-quantized heavy-hole subband H2 with nonmonotonic dispersion law \cite{DyakonovJETP-1982-28,GerchikovPSSB-1990-29} (see Fig.~\ref{fig1:BandStrucHgTe}).

A substantial overlap about 6.45meV of the valence H2 and conduction H1 subbands is obtained when the strain is considered in calculations, but this overlap would not be felt experimentally in a single QW at electron densities $n_s \gtrsim 1.5 \times 10^{15}$m$^{-2}$  since the Fermi level $E_F$ is above the overlap region.

In our system the conduction is carried out by electrons of the H1 subband with a small value of the effective mass at $E_F$: $m_{eff}=(0.022-0.026)m_0$  for $n_s = 1.5 \times 10^{15}$m$^{-2}$~ \cite{OlshanetskyJETPL2006-22,KvonLTP-2009-30,YaPE-2010-31,ZhangPRB-2001-32} and with a large value of $g$-factor, $g \cong 50$ \cite{YaPE-2010-31}.

\subsection{\label{sec:22}The Landau level fan diagram}

\begin{figure}[b]
\includegraphics[width=90mm]{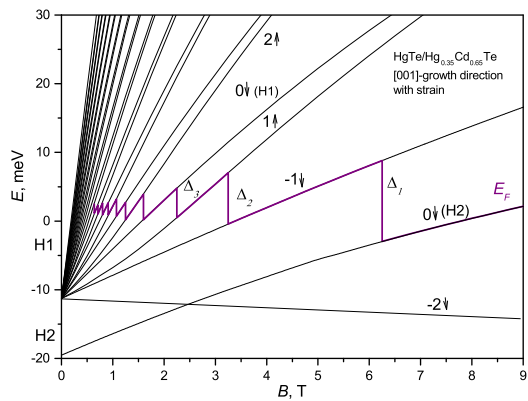}
\caption{\label{fig2:LLHgTe} Landau levels of  H1 and H2 subbands for an $n$-type HgTe/Hg$_{0.3}$Cd$_{0.7}$Te (001) QW as a function of magnetic field for $d_{QW}= 20$nm, $n_s =1.5 \times 10^{15}$m$^{-2}$. The Landau levels are labeled in accordance with the notations of Ref.~\onlinecite{NovikPRB-2005-27}: the quantum numbers $N = -2,-1,0,…,$ and the arrows $(\uparrow,\downarrow)$ indicate the dominant spin orientation of the state. The thick line represents $E_F$.}
\end{figure}

Calculated Landau level (LL) spectrum of a HgTe/Hg$_{0.35}$Cd$_{0.65}$Te ([001]) QW is shown on Fig.~\ref{fig2:LLHgTe} for the structure whose subband dispersion is presented in Fig.~\ref{fig1:BandStrucHgTe}. The LL notation corresponds to the notations of Ref.~\onlinecite{NovikPRB-2005-27}.   
 
It is seen from Fig.~\ref{fig2:LLHgTe} that Landau levels are essentially nonequidistant and nonlinearly depend on magnetic field that is caused by mixed nature of the H1 and H2 subbands in the inverted-band regime due to a coupling between heavy-hole to light-particle states at finite in-plane wave vectors $k_{||}$. Only the lowest Landau level ($N = -2\downarrow$) of the H1 subband contains pure heavy-hole states, which do not mix with the light-particle states \cite{NovikPRB-2005-27} (see also Refs.~\onlinecite{ArapovJETPL-1994-33,YakuninNanotechnol-2000-34} and references therein). 

In Refs.~\onlinecite{ArapovJETPL-1994-33,YakuninNanotechnol-2000-34} it is shown that this level is of the same nature in two and in three dimensions and it is mapped on the $b$ set level of Guldner et al. \cite{GuldnerPRB-1973-35} 
$$E_b(0)=\left(\frac{e \hbar B}{m}\right)\cdot \varepsilon_b(0) \mbox{, where}$$
\begin{equation}
\varepsilon_b(0)=\frac{1}{2}(\gamma_1+\overline{\gamma})-\frac{3}{2}\kappa.
\label{eq:1}
\end{equation}
For the set of $\Gamma_8$ Luttinger parameters for HgTe ($\gamma_1=12.8$, $\overline{\gamma} =8.4$ and $\kappa =10.5$ \cite{GuldnerPRB-1973-35}) we have:
\begin{equation}
3\kappa>(\gamma_1+\overline{\gamma}).
\label{eq:2}
\end{equation}
and, according to (\ref{eq:1}), the level $N = -2\downarrow$ {\slshape lowers} its energy linearly with increasing magnetic field and thus reveals its {\slshape hole-like} character. All the other Landau levels of the H1 subband show an {\slshape electron-like} character: they {\slshape rise} in energy with magnetic field due to the coupling with light-particle states. 

It is also seen from Fig.~\ref{fig2:LLHgTe} that the Landau level of the H2 subband with $N=0\downarrow$ becomes the highest H2 LL at $B \gtrsim 5$T as a result of mixing between the heavy and light states \cite{AncilottoPRB-1988-36}. The unusual behavior of the $N = -2\downarrow$ level from the conduction subband H1 in inverted band HgTe QW together with the peculiar dispersion of the $N=0\downarrow$ level from the topmost valence subband H2 leads to the crossing of conduction- and valence-subband states at a some  value $B_c$ of the magnetic field (see Fig.~\ref{fig2:LLHgTe}).

Such a behavior is specific for HgTe QW and has been examined theoretically and experimentally (see, for example, Ref.~\onlinecite{SchultzPRB-1998-37}). In our case the intersection point is immersed in the thick of valence subband Landau levels  and is irrelevant for our measurements, while just the $N=0\downarrow$ level of the H2 subband appears as the lowest Landau level of the conduction band at $B\gg B_c$.

Our further plan is as follows.  First, to determine the activation energies in $\rho_{xx}(T)$ in the QHE regime for integer filling factors: $i = 1$ (the gap between $N=0\downarrow$ LL of H2 subband and $N=-1\downarrow$ LL of the H1 subband), $i = 2$ (gap $N=-1\downarrow \to N=1\uparrow$ of the H1 subband) and $i = 3$ (gap  $N=1\uparrow \to N=0\downarrow$ of the H1 subband). Then to investigate the low-temperature variable-range hopping transport at $i = 1$ and $i = 2$ QHE  plateau regions and, finally, to analyze the temperature dependence of the $1 \to 2$ QHE plateau-plateau transition width.

\section{\label{sec:3}Experimental results and discussions}

\begin{figure}[b]
\includegraphics[width=90mm]{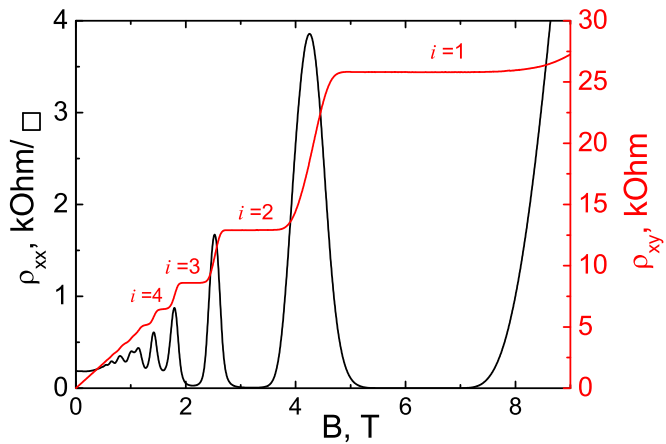}
\caption{\label{fig3:roXXXY} The longitudinal and Hall components of the magnetoresistivity tensor as functions of magnetic field $B$ at $T = 2.9$K.}
\end{figure}

Figure~\ref{fig3:roXXXY} shows the magnetic-field dependences of the longitudinal $\rho_{xx}$ and Hall $\rho_{xy}$  resistivities for the sample under study at $T = 2.9$K. We observe the features characteristic of the QHE regime, i.e., the regions of plateaus in the $\rho_{xy}(B)$ dependences  ($\rho_{xy} = h/ie^2$) with rather sharp transitions between them: for $B \geq 2$T, we can see plateaus with numbers $i = 4, 3, 2, 1$. 

\subsection{\label{sec:31}Activation energies}

We can extract the activation gaps  between adjacent LLs from the temperature dependence of the minima in $\sigma_{xx}$ with a Fermi-Dirac fit and compare the results with theoretical calculations of the Landau level dispersions shown in Fig.~\ref{fig2:LLHgTe}. The activation energy achieves its maximum value, $E_A^{max}$ , at an integer value of the filling factor $\nu$. The mobility gap width estimated as $\Delta = 2E_A^{max}$  is closely related to the energy separation between adjacent LLs:  $\Delta \cong |E_N - E_{N'}|$.

\begin{figure}[b]
\includegraphics[width=90mm]{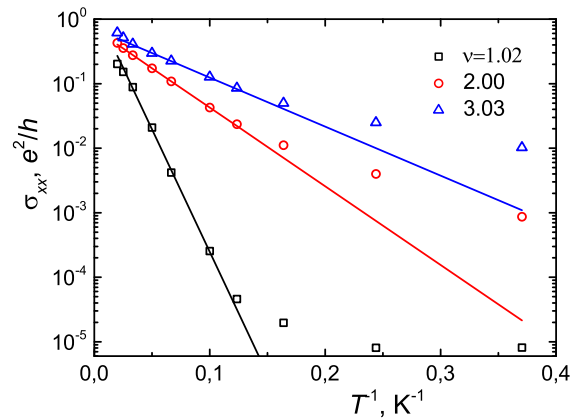}
\caption{\label{fig4:sXX} $\sigma_{xx}(T)$ dependencies at $\nu = 1$, $\nu = 2$  and $\nu = 3$ with a fitting  by Arrhenius equation  (straight lines).}
\end{figure}

Fig. ~\ref{fig4:sXX} shows a fit of  $\sigma_{xx}(T)$ dependencies for investigated sample at $\nu = 1$, $\nu = 2$ and $\nu = 3$ by the Arrhenius equation  (straight lines in figure) in the range of more than one (for  $\nu = 2$ and 3) or even  three (for $\nu = 1$) orders of conductivity  at  $T \cong (10 -50)$K. Deviations of experimental points from straight lines for $T \lesssim 10$K  are explained by the variable range hopping among localized states at $E_F$, which usually dominates for sufficiently low $T$ (see section \ref{sec:32}).

\begin{table}[h]
\caption{\label{tab1:Par} The values of magnetic fields  $B_\nu$, of the experimentally obtained activation energies $E_A$, of corresponding energy gaps  $\Delta_\nu^{exp}= 2 E_A$, and of calculated energy gaps $\Delta_\nu^{teor}$  at LL filling factors $\nu  = 1$, 2 and 3.}
\begin{tabular}{|l|c|c|c|}
\hline
~Parameters~ & ~$\nu=1$~ & ~$\nu=2$~ & ~$\nu=3$~ \\
\hline
~$B_\nu$, T & 6.3 & 3.15 & 2.1 \\
\hline
~$E_A$, meV & 6 & 2.4 & 1.8 \\
\hline
~$\Delta_\nu^{teor}$, meV & 11.6 & 6.9 & 3.8 \\
\hline
~$\Delta_\nu^{exp}$, meV & 12 & 4.8 & 3.6 \\
\hline
~$\Delta_\nu^{exp}$, K & 139.2 & 55.7 & 41.8 \\
\hline
\end{tabular}
\end{table}

Table~\ref{tab1:Par} presents a comparison of the extracted activation energies with theoretical calculations from Fig.~\ref{fig2:LLHgTe}. It is seen that the experimentally and theoretically obtained energy gaps are in reasonably good agreement and thus the behavior of the sample is well described by our $\mathbf k \times \mathbf p$ model. In particular, we don't observe an explicit manifestation of a strong Rashba spin-orbit splitting caused by an asymmetry of QW confinement potential (see Fig. 1c in Ref.~\onlinecite{NovikPRB-2005-27}).

In HgTe-based 2D structures, the activation energies were determined earlier \cite{KozlovAPL2014} from the temperature dependences of the longitudinal resistivity in the regions of quantized Hall plateaus for the filling factors $\nu$ of 1 and 2 in a 6.6 nm HgTe quantum well at magnetic fields up to 34 T at nitrogen temperatures. The indications of the large values of the $g$-factor (about 30-40) were found.

In Ref.~\onlinecite{KhouriPRB-2016-24} the QHE in HgTe QWs with a finite band gap below and above the critical thickness $d_c$ ($d_{QW} = 5.9$nm and 11nm) has been studied up to temperatures of about 50K. They extracted energy gaps between LLs of $(40 - 45)$meV for $\nu = 1$ and $\sim 25$meV for $\nu = 2$ from the temperature dependent magnetotransport measurements, in good agreement with the Landau level spectrum obtained from  calculations.

\subsection{\label{sec:32}Hopping conductivity}

In this section we discuss a conduction process within the ranges of quantum Hall plateaus. 
D.G. Polyakov, B.I. Shklovskii and I.L. Aleiner \cite{PolyakovPRL-1993-38,PolyakovPRB-1993-39,AleinerPRB-1994-40} showed that in the strongly localized electron system in the QHE-plateau regions, the dominant mechanism of the low-temperature transport must be the variable-range hopping (VRH) near $E_F$ (see also Ref. \onlinecite{Shklovskii-1979-41,Shklovskii-1984-42}). Consequently, the temperature induced conductivity far from a QHE peak should be exponentially small. The exponential factor should grow as $E_F$ approaches the LL center due to the divergence of the localization length:
\begin{equation}
\xi \propto |B-B_N|^{-\gamma} \propto |\nu-\nu_c|^{-\gamma},
\label{eq:3}
\end{equation}
where $B_N$ is the value of $B$ at which $E_F$ is in the center of $N$-th LL and the critical filling factor $\nu_c$ is a half integer value of $\nu$. Here $\gamma$ is the critical exponent. The analytical calculation of $\gamma$ is a difficult problem; for the short-ranged impurity potential, numerical methods give $\gamma~=~2.35~\pm~0.03$ (see, for example, reviews \cite{KramerPR-2005-43,HuckesteinRMP-1995-44}).

For noninteracting 2D electrons, the VRH Mott's law gives \cite{MottJNCS-1968-45}:
\begin{equation}
\sigma_{xx} \sim \frac{1}{T} \exp\left[-\left(\frac{T}{T_M}\right)^\frac{1}{3}\right]
\label{eq:4}
\end{equation}
with $kT_M= \beta /\xi^2g(\varepsilon_F)$, where $g(\varepsilon_F)$ is the finite density of states at $E_F$ and numerical constant $\beta =13.8\pm 0.8$ \cite{SkalFTT-1976-46}.  

However, in the QHE regime screening is poor and Coulomb repulsion must be included. This is the Efros - Shklovskii (E-S) VRH regime \cite{EfrosJPC-1975-47}, where the 2D density of states $g(\varepsilon) \sim |\varepsilon - \varepsilon_F|$ yields 
\begin{equation}
\sigma_{xx} \sim \frac{1}{T} \exp\left[-\left(\frac{T}{T_0}\right)^\frac{1}{2}\right]
\label{eq:5}
\end{equation}
with  $kT_0 = Ce^2/4\pi \epsilon \epsilon_0\xi$, determined by the Coulomb energy on the localization length $\xi$, $C \cong 6.2$ is a numerical constant, $\epsilon $ - dielectric constant. 

Measuring $T_0(\nu)$ allows to determine $\xi$ and to probe the scaling law: 
\begin{equation}
T_0 \sim \frac{1}{\xi(\nu)} \sim |\nu-\nu_c|^\gamma.
\label{eq:6}
\end{equation}

The concept of VRH conduction proved to be very productive for the interpretation of thermally activated transport in the plateau regions of the integer QHE. Direct determination of $\xi$ and its scaling exponent from the E-S's VRH was done in previous measurements performed on conventional 2DEGs, including Si-MOSFETs,  GaAs/AlGaAs  and n- InGaAs/InAlAs heterojunctions, as well as in the monolayer graphene and in other graphene-based low-dimensional structures (see a comprehensive list of references in Ref.~\onlinecite{ArapovJLTP2016-25,GudinaPSSC2016-26}). 

It is worth to highlight the thorough work \cite{FurlanPRB-1998} where for GaAs/AlGaAs heterostructures the divergence of localization length with exponent $\gamma =2.3$ was estimated. The results \cite{BennaceurPRB-2012-48} for graphene monolayers should be emphasized: it is the first observation of the crossover from the Efros-Shklovskii's to Mott's VRH for a quantum Hall system, which happens when the localization length exceeds the screening length set by the metallic gate in accordance with the Aleiner and Shklovskii prediction \cite{AleinerPRB-1994-40}.

For HgTe QW, the first study of the temperature-induced transport at the QHE resistivity minima corresponding to the QHE-plateau regions was done within the concept of variable-range hopping conductivity in our previous works \cite{ArapovJLTP2016-25,GudinaPSSC2016-26} on the HgTe/HgCdTe system with inverted band structure. Here we revisited this theme summarizing the results obtained and presenting the data in a more universal form.

\begin{figure}[b]
\includegraphics[width=90mm]{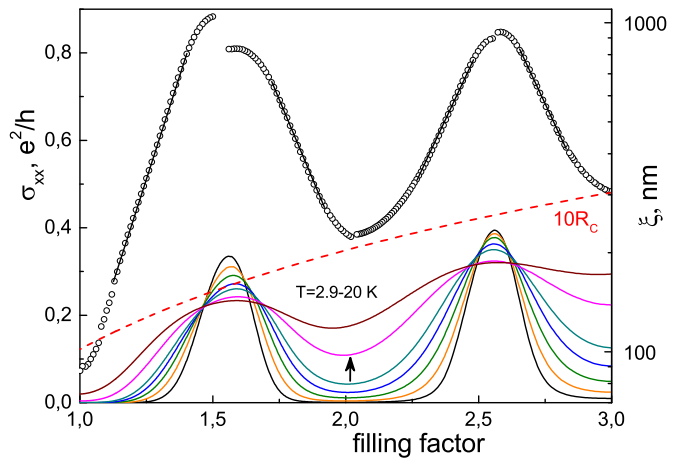}
\caption{\label{fig5:sXX-Rc} Conductivity $\sigma_{xx}$ in the QHE regime at different temperatures and the localization length $\xi$ extracted from E-S VRH fits of $\sigma_{xx}(T)$ in comparison with  cyclotron radius $R_C$ as functions of a filling factor.} 
\end{figure}

\begin{figure}[b]
\includegraphics[width=90mm]{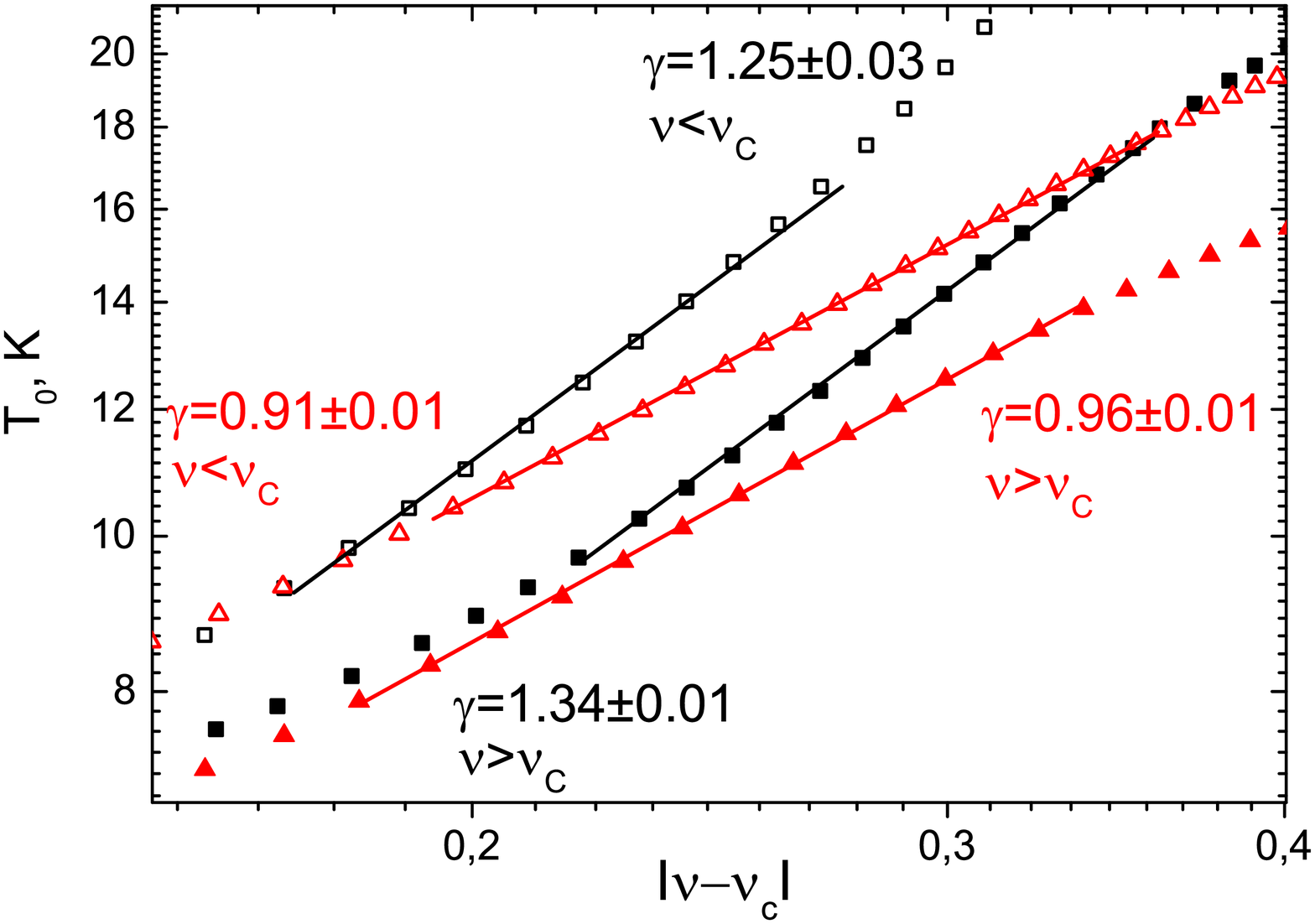}
\caption{\label{fig6:T0} Characteristic temperature $T_0$ as a function of $|\Delta \nu |=|\nu-\nu_c|$ at $\nu >(<) \nu_c$  (filled (open) symbols) for $\nu_c = 1.53$ (squares) or 2.56(triangles).} 
\end{figure}

\begin{figure}[t]
\includegraphics[width=90mm]{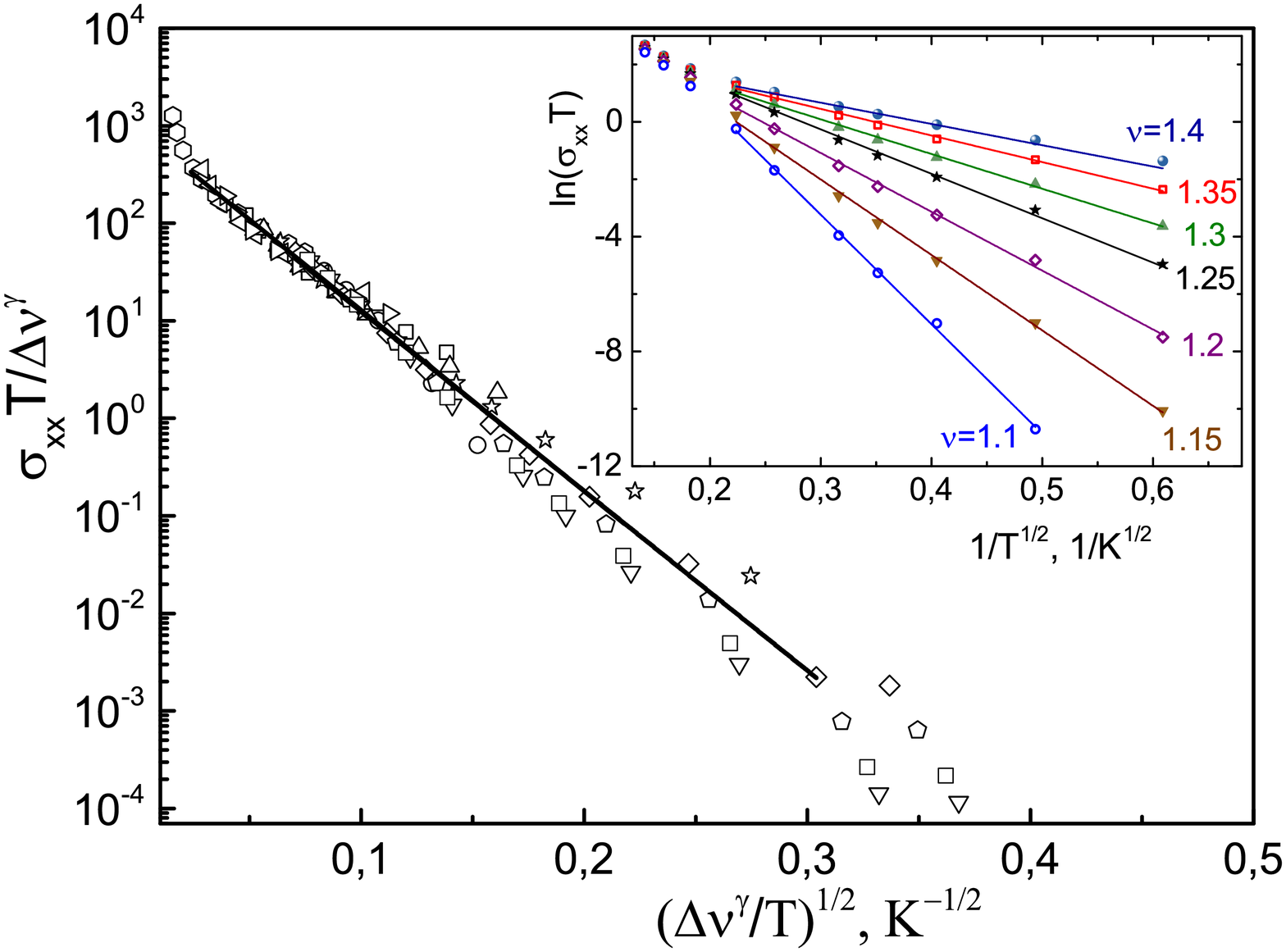}
\caption{\label{fig7:S-x} Conductivity $\sigma_{xx}(T,\nu)$ as a function of the scale parameter $x = |\Delta \nu|^\gamma/T$
at $\nu < \nu_c$  for  $\nu_c = 1.53$ ($T = 2.9 - 10$K, $|\Delta \nu|< 0.35$). \\
 Inset: the log plot of $T\sigma_{xx}(T)$ as a function of $1/T^{1/2}$. Solid lines are a E-S's law (Eq.~\ref{eq:5}) fit of the data.}
\end{figure}

The longitudinal conductivity $\sigma_{xx}$ as a function of a filling factor at actual QHE  minima  near  $\nu = 1, 2$ and 3 is presented in Fig.~\ref{fig5:sXX-Rc}. The activation behavior of $T\sigma_{xx}(T)$ on $(1/T)^{1/2}$ in minima with $\nu =1$ can be seen in the inset of Fig.~\ref{fig7:S-x}. Solid lines are the E-S's law (Eq.~\ref{eq:5}) fit of the data with the temperature $T_0$ (Fig.~\ref{fig6:T0}) obtained from the fitting.

In Fig.~\ref{fig5:sXX-Rc} we also show the localization length $\xi(\nu)$ computed by (Eq.~\ref{eq:6}) from the values of $T_0(\nu)$, extracted from E-S VRH fits of $\sigma_{xx}(T)$, for continuous values of the filling factor. 

For comparison, a graph of cyclotron radius, $R_C$ (multiplied by ten) dependence on $\nu$ for different magnetic fields is also shown. The minimal localization length found in the middle of QH plateaus is $\xi_{min} \sim 100$nm for $\nu = 1$ ($B_1 = 6.3$T) and $\xi_{min} \sim 200$nm for $\nu = 2$ ($B_2 = 3.15$T) and $\nu = 3$ ($B_3 = 2.1$T), which is about ten times larger than $R_C$. 

We emphasize that $\xi_{min}(\nu)$ dependence is in correlation just with $R_C(\nu)$ dependence (but not with $l_B(\nu)$) in accordance with conclusions of Fogler et. al \cite{FoglerPRB-1998-49}. The fact that $\xi_{min}\gg R_C$ indicates a large-scale character of the random impurity potential.

Fig.~\ref{fig6:T0} shows a characteristic temperature $T_0$ versus $|\Delta \nu|=|\nu-\nu_c|$ at $\nu >(<) \nu_c$ for $\nu_c \approx 1.5$ and $\approx 2.5$. The linear variation indicates regions of $|\Delta \nu|$ where the corresponding $\gamma$ is a reasonable exponent:  
at $\nu_c = 1.53~~~ \gamma \approx 1.3 ~ (0.2 \lesssim |\Delta \nu|  \lesssim 0.33)$;  at $\nu_c = 2.56~~~ \gamma \approx 0.93 ~ (0.2 \lesssim |\Delta \nu|  \lesssim 0.35)$.  

To probe the universality of a scaling law in the surrounding area of  $\nu_c = 1.5$ or 2.5 we have plotted all the conductivity data $\sigma_{xx}(\nu , T)$  as a function of a single parameter $x = \frac{|\Delta \nu|}{T} \sim \frac{T_0}{T}$. The values for $\gamma$ are taken from the power-law fits of $T_0$ depending on $|\nu - \nu_c|$ (see Fig.~\ref{fig6:T0}). An example of scaling behavior  for conductivity $\sigma_{xx}$ as a function of $x$ at $\nu_c \approx 1.5$  is shown in Fig.~\ref{fig7:S-x}.

Rescaling the axes  shows that all experimental points fall onto straight lines for  nearly four orders of conductivity  at $|\Delta \nu| \lesssim 0.35$ both  for  $\nu_c = 1.5$ and 2.5 . Thus, Figs~\ref{fig6:T0} and \ref{fig7:S-x} demonstrate the accurate enough universal single parameter scaling  for the conductivity in the VRH regime at QHE plateaus in analogy with the observations of scaling behavior for VRH in GaAs/Al$_x$Ga$_{1-x}$As structure \cite{TaoPRL-2007-50,ZhaoPRB-2008-51}.

 Note that the exponent values: $\gamma = 1.31 \pm 0.03 ~ (0.2\lesssim |\Delta \nu| \lesssim 0.33)$ at  $\nu_c = 1.53$  and   $\gamma = 0.93 \pm 0.03 ~(0.2\lesssim |\Delta \nu| \lesssim 0.35)$ at  $\nu_c = 2.56$ -  correlate well with  the exact theoretical result $\gamma= \frac{4}{3}$,  obtained in a theory of classical percolation for systems with large-scale impurity potentials \cite{TrugmanPRB-1983-52} (see discussion in section \ref{sec:42}).

\subsection{\label{sec:33}Plateau - plateau transition }

The integer quantum Hall effect regime can be considered as a sequence of insulator-metal-insulator quantum phase transitions when the density of states of a disordered 2D system is scanned by $E_F$ in a quantizing magnetic field. We analyze data on the magnetic-field and temperature dependences of conductivity in the regions of the  plateau-plateau transitions within framework of a scaling hypothesis for a quantum phase transition \cite{HuckesteinRMP-1995-44}. 

\begin{figure}[b]
\includegraphics[width=90mm]{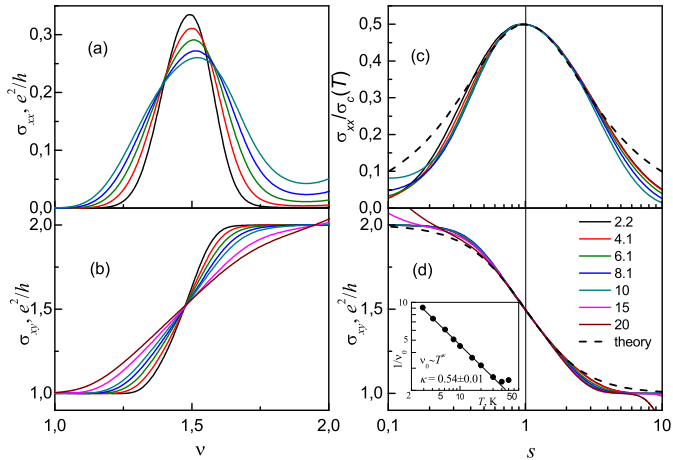}
\caption{\label{fig8:Sig-s} Longitudinal $\sigma_{xx}$ and Hall $\sigma_{xy}$ conductivities as  functions (a, b) of the filling factor $\nu$   or (c, d) of the scaling variable $s$  in the PPT $1\to 2$ region at  $\nu_c = 1.5$. The dashed  black lines indicate  fitting to Eq.\ref{eq:7}. Inset to (d) shows temperature dependence of $1/\nu_0$ for the $1 \to 2$ transition. }
\end{figure}

Let us concentrate our investigation on the region of the transition between the first and second QHE plateaus and analyze the temperature dependence of the transition width in the vicinity of the critical magnetic field ($B_c = 4.1$T). Longitudinal $\sigma_{xx}$ and Hall $\sigma_{xy}$ conductivities as  functions of the filling factor $\nu$  calculated by experimental data on $\rho_{xx}$ and $\rho_{xy}$ in the range $1 < \nu < 2$ with the critical value $\nu_c = 1.5$ is shown in  of Figs~\ref{fig8:Sig-s}a,b.

For a treatment of the data in a region of $(i-1) \to i$ PPT, we used interpolation formulas via the so-called scattering parameter (see, e.g., Ref.~\onlinecite{ColeridgePRB-1999-53} and references therein):
\begin{equation}
\sigma_{xx}=\frac{s}{1+s^2}, ~~~
\sigma_{xy}=i-\frac{s^2}{1+s^2}
\label{eq:7}
\end{equation}
where $s$ varied from 0 to $\infty$, it is unity at the critical point $\nu = \nu_c$ and exponentially depends on the filling factor in the vicinity of the critical point:
\begin{equation}
s=\exp \left( -\frac{\Delta \nu}{\nu_0(T)}\right)
\label{eq:8}
\end{equation}
Here $\nu_0(T)$ is the effective bandwidth of delocalized states at the temperature $T$.

The inset of Fig.~\ref{fig8:Sig-s}d shows the log-log temperature dependence of $1/\nu_0$ for the $1 \to 2$ transition in the sample under study. We can see that the dependence $\nu_0(T)$ is described by a straight line with good accuracy in a wide temperature range $2.9 \le T \le 30$K. Thus, the temperature behavior of the transition width is defined by the scaling dependence $\nu_0 \sim T^\kappa$  with the critical exponent $\kappa = 0.54 \pm 0.01$. 
This value does not correspond to the "classical" result $\kappa \approx 0.42$ for the short-range
scattering potential , however, it correlates well with experimental results for systems with large-scale impurity potentials (see discussion in section \ref{sec:42}).

Test of the scaling with a single parameter $s \sim |\nu - \nu_c|/ T^\kappa$  with $\kappa = 0.54$  for the dependences both of $\sigma_{xx}(\nu, T)$ and of $\sigma_{xy}(\nu, T)$ at $1 \to 2$ QHE transition is presented on Fig.~\ref{fig8:Sig-s}c,d. It is seen that a scaling behavior is valid for $0.7 \lesssim s\lesssim 2.5$ at $T = (2.9 - 10)$K for $\sigma_{xx}$ and $T = (2.9 - 20)$K for $\sigma_{xy}$.

\section{\label{sec:4}Discussion of results}

The plateau-plateau transition (between neighboring quantum Hall liquids through an intermediate metal phase) was considered as an electron localization-delocalization-localization quantum phase transition already in the first papers on QHE interpretation \cite{AokiPRL-1985-54,PruiskenPRL-1988-55} and is widely treated at present within the framework of a scaling hypothesis (see, e.g., the reviews \cite{KramerPR-2005-43,HuckesteinRMP-1995-44,KramerPE-2003-56,GantmakherPU-2008-57,EversRMP-2008-58}).

The scaling hypothesis is based on a concept that at the absolute zero of temperature the localization length diverges at the critical energy $E = E_c$ of the phase transition at the center of the broadened Landau level with a universal exponent $\gamma$ (the critical exponent of the localization length) \cite{HuckesteinRMP-1995-44,HuckesteinPRL-1990-59}:
\begin{equation}
\xi(E)=\frac{\xi_N}{(E-E_c)^\gamma}
\label{eq:9}
\end{equation}
where the constant $\xi_N$ depends on microscopic details of the randomness and
on the Landau band index $N$. For a short-range random potential $\xi_N$ is of the order of cyclotron radius $R_C$\cite{FoglerPRB-1998-49}.

\begin{figure}[b]
\includegraphics[width=90mm]{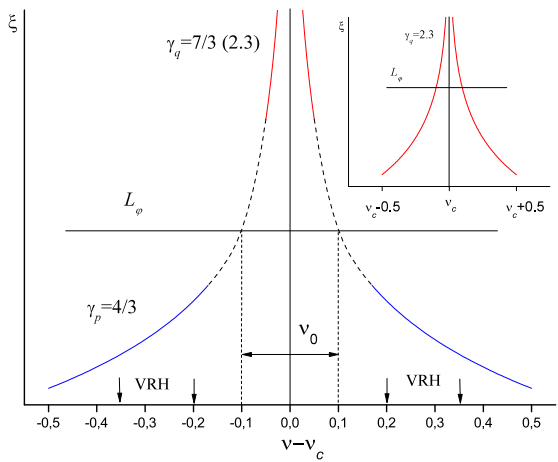}
\caption{\label{fig9:Xi-nu} Localization length  dependencies on  the filling factor $\nu$ within a modern theoretical conception for a large-scale impurity potential in QHE regime (see a description in the text). Inset:  $\xi(\nu)$ dependencies for a short-range impurity potential. A critical exponent of localization length theoretically is $\gamma \approx 2.3$ for
$|\nu-\nu_c| \le   0.5$.
}
\end{figure}

At finite temperatures, the region of delocalized states at the Landau level center can be described by an energy range where the localization length $\xi(E)$ increases to a characteristic length $\xi(E) > L_\varphi$  (Fig.~\ref{fig9:Xi-nu}). Here $L_\varphi \sim T^{-p/2}$ is the phase coherence length  and the dynamical exponent $p$ depends on the inelastic scattering mechanism. 

At $\xi(E) < L_\varphi$  electronic states remain localized and the bandwidth $\nu_0$ of delocalized states is determined from the condition $\xi(E) \cong L_\varphi$ \cite{HuckesteinRMP-1995-44,AokiPRL-1985-54,PruiskenPRL-1988-55}. Thus the width of the transition between neighboring QHE plateaus, as well as the width of the corresponding peak in the magnetic-field dependence $\sigma_{xx}(B)$ should tend to zero by the power-law dependence $T^\kappa$, where $\kappa = p/2\gamma$.

\subsection{\label{sec:41}Short-range random potential }

The theoretical investigations of the critical behavior of noninteracting electrons in the quantum Hall system with the short-ranged disorder potentials led to the conclusion about a single diverging length scale and the results of extensive efforts  on numerical simulations for the critical exponent gave the value $\gamma = 2.35 \pm 0.03$ (see, for example, reviews \cite{KramerPR-2005-43,HuckesteinRMP-1995-44} and the detailed table in the review \cite{KramerPE-2003-56}).

The critical exponent $\kappa = 0.42$ experimentally determined for the first time in the classical study \cite{WeiPRL-1988-60} for InGaAs/InP systems ($\kappa = 0.42 \pm  0.04$) is in excellent agreement with the conclusions of new unique studies of Al$_x$Ga$_{1 - x}$As /Al$_{0.33}$Ga$_{0.67}$As systems in the region of alloy scattering ($\kappa = 0.42 \pm  0.01$) \cite{LiPRL-2005-61} and with the results of recent studies of the first and second Landau levels (both for electrons and holes) in single layer graphene \cite{BennaceurPRB-2012-48,GiesbersPR-2009-62}.

The observable exponent  $\kappa = 0.42$ is compatible with a numerical short-ranged potential value $\gamma \approx 2.3$ for  the Fermi-liquid dynamical  exponent  $p=2$  as  it is believed to be the case by Li et al. \cite{LiPRL-2005-61} along with the pioneering work of Wei et al. \cite{WeiPRL-1988-60}.

Although the value of the parameter $\kappa$ is currently the subject of discussion, there is a consensus that $\kappa = 0.42$ indeed describes transitions in the QHE regime (when they are not masked by macroscopic inhomogeneities) for systems with short-range scattering potentials \cite{PruiskenIJMP-2010-63,BurmistrovAP-2011-64}.

\subsection{\label{sec:42}Large-scale random potential}

However, in sharp contrast to the short range alloy potential scattering in InGaAs/InP samples \cite{WeiPRL-1988-60}, classical and most studied AlGaAs/GaAs heterostructure has long range Coulomb scattering on remote (by a spacer) ionized impurities which results in nonuniversality of the temperature exponent $\kappa$ (see both the early \cite{KochPRL-1991-65,WeiPRB-1992-66,YooSSC-1994-67,KochSST-1995-68} and the recent works \cite{TaoPRL-2007-50,ZhaoPRB-2008-51,HohlsPRL-2002-69,DodooJPCM-2014-70,Nakajima-2007-71,LiPRB-2010-72,XuebinPRB-2016-73}).

In  modulation-doped GaAs/AlGaAs heterostructures the values $\kappa > 0.42$ are regularly observed (see Table ~\ref{tab2:Kappa} in Appendix).
In Table 2 the results for critical exponent ($\kappa$) values in  modulation-doped GaAs/AlGaAs heterostructures from the works of the years 1991-2016 have been collected, and the  "nonuniversal"  values of parameter $\kappa$ in the range of ($0.5 -0.75$) come to  light.

The fact that a slowly varying potential turned out to be the generic type of disorder in the standard AlGaAs/GaAs heterostructure has led historically to semiclassical considerations (percolation picture) of delocalization near the Landau band center. The ideas, which relate localization to the classical percolation in the context of the integer quantum Hall effect, have been developed intensively by a number of authors (see the article of Prange \cite{Prange-1987-74} for exhaustive information).

In the theoretical calculations, an exponent $\gamma = 4/3$ was obtained within a model of classical percolation \cite{TrugmanPRB-1983-52,LeePRL-1993-75}. On the other hand, after including the effect of quantum tunneling, the universal critical exponent $\gamma = 7/3$ results from a model of quantum percolation \cite{LeePRL-1993-75,MilnikovJETPL-1989-76} (see a clear exposition of arguments in a number of reviews \cite{KramerPR-2005-43,HansenCM-1997-77,Burmistrov-2015-78}).

The percolation model for QHE supplemented by quantum effects \cite{LeePRL-1993-75,MilnikovJETPL-1989-76} provides a physical background for the Chalker-Coddington network model \cite{ChalkerJPC-1988-79} - a generic model, which is assumed to describe the \emph{universal} quantum mechanical properties of non-interacting electrons in two dimensions in the presence of a random potential subject to a strong perpendicular magnetic field. An overview of the random network model, invented by Chalker and Coddington, and its generalizations is provided, for example, in \cite{KramerPR-2005-43}.

In a seminal paper on percolation and quantum tunneling in the integer quantum
Hall effect \cite{ChalkerJPC-1988-79} a network model for localization in the QHE regime has been introduced that made it possible to numerically simulate a system where the disorder potential varies slowly on the magnetic length scale. Using the simplifying features of a slowly varying potential in the model the quantum tunneling and interference effects were incorporated. It turned out that the network model contains the features necessary for a qualitative understanding of the integer quantum Hall effect: localized states in the Landau band tails and extended states in the band center, existing only at one energy. To this extent, the classical picture survives the introduction of quantum tunneling.

There are, however, quantitative changes. In the classical picture, as was shown
earlier \cite{TrugmanPRB-1983-52,LeePRL-1993-75}, the localization length diverges with an exponent $\gamma = 4/3$. For the
network model \cite{ChalkerJPC-1988-79} the value $\gamma = 2.5 \pm 0.5$ was found in a reasonable agreement with estimates for a rapidly varying potential.

The modern theoretical network models for the large-scale impurity potential with the quantum tunneling give a numerical value of the critical exponent $\gamma \approx 2.3$ in the immediate vicinity of the critical energy $E = E_c$ ($\nu = \nu_c$) (see \cite{KramerPR-2005-43,HuckesteinRMP-1995-44} and references therein) in accordance with the findings of Ref.~\onlinecite{ChalkerJPC-1988-79}. On the other hand, far from the critical energy, dependence of $\xi$ on $E - E_c$ (on $\nu = \nu_c$) is determined by the model of classical percolation with $\gamma = 4/3$ (see Fig.~\ref{fig9:Xi-nu}).

\subsection{\label{sec:43}Interpretation of our results }

A schematic representation of localization length divergences with $|\nu-\nu_c|$:
\begin{equation}
\xi(\nu)=|\nu-\nu_c|^{-\gamma}
\label{eq:10}
\end{equation}
is provided in Fig.~\ref{fig9:Xi-nu} for a short-range (inset) and a large-scale impurity potentials according to theoretical considerations described above.

The solid lines in Fig.~\ref{fig9:Xi-nu} are: the dependence (\ref{eq:10}) with $\gamma = \gamma_P =  4/3$ in  regions of classical percolation (blue lines) and with $\gamma = \gamma_q$   in regions of the  quantum-tunnelling processes  (red lines). Here  $\gamma_q = 7/3$ within a modified  percolation model \cite{MilnikovJETPL-1989-76} and $\gamma_q = 2.3$ within the modern network model \cite{KramerPR-2005-43,HuckesteinRMP-1995-44}.

The dash lines in Fig.~\ref{fig9:Xi-nu} show an intermediate region of Eq. (\ref{eq:10}) with $4/3< \gamma < 7/3$ (or 2.3) that gives  $0.42< \kappa < 0.75$ (with the exponent $p=2$) in the interspace of crossover from a classical percolation to the quantum tunneling as pointed out by Li et.al \cite{LiPRL-2005-61}.

We believe that the critical exponent value for the bandwidth of delocalized states, $\kappa = 0.54 \pm 0.01$ obtained by us, as well as a number of results with $\kappa = 0.5 - 0.75$ for systems with large-scale impurity potentials 
(see Table 2 in Appendix)  are driven by a situation schematically represented on Fig.~\ref{fig9:Xi-nu}: the line $L_\varphi =const$ crosses the curves $\xi(\nu)$ just at the intermediate region of $\gamma$ values. This situation, quite possibly, is typical for  modulation-doped GaAs/AlGaAs heterostructures \cite{TaoPRL-2007-50,ZhaoPRB-2008-51,HohlsPRL-2002-69,DodooJPCM-2014-70,Nakajima-2007-71,LiPRB-2010-72,XuebinPRB-2016-73} resulting in "nonuniversal"  values of parameter $\kappa$ in the range of ($0.5 - 0.7$).
Note that for HgTe based heterostructure with inverted band spectrum ($d_W = 21$nm, $n = 1.5\times 10^{15}$m$^{-2}$) a scaling regime in the QHE has been investigated earlier by Olshanetsky et al. in \cite{OlshanetskyJETPL2006-22} and a lower value, $\gamma = 0.49$, was obtained for $1 \to 2$ PPT at helium temperatures ($0.3 - 3.0$)K.

On the other hand, in a recent study \cite{KhouriPRB-2016-24} on HgTe quantum well with $d_W = 5.9$nm, that is below critical thickness $d_c$, Khouri et.al  observed a quantized Hall conductivity up to 60K at high carrier concentration $n = 4.6\times 10^{15}$m$^{-2}$. From the scaling behavior, realized for the PPT in a wide temperature range ($0.3 - 30$)K, they have found the coefficient $\kappa = 0.45 \pm 0.04$ for the transition $\nu = 2 \to 1$ and $\kappa = 0.40 \pm 0.02$ for the $\nu = 3 \to 2$  transition  in excellent agreement with the universal scaling theory for systems described by short-range scattering (see, for example, Ref.~\onlinecite{HuckesteinRMP-1995-44}).

The high carrier concentration\cite{KhouriPRB-2016-24}, achieved  by applying gate voltages for a tuning the Fermi energy deep into the conduction band, apparently promotes an effective screening of large-scale potential fluctuations as well as of any inhomogeneities.

As for the variable-range hopping conduction, this mechanism of the low-temperature transport takes place in the regions of localized states at $\xi(E) < L_\varphi$. In these regions, localization length $\xi$ and its critical exponent may be determined by a direct way.

For a short-range potential (see inset on Fig.~\ref{fig9:Xi-nu}) a combination of the values of $\kappa \approx 0.42$ (if $p = 2$) from an analysis of PPT width at $\xi > L_\varphi$ and $\gamma \approx 2.3$ from VRH analysis at $\xi < L_\varphi$ should be observed.Just such a situation is, for example, implemented for monolayer graphene \cite{BennaceurPRB-2012-48,GiesbersPR-2009-62} (except the zero Landau level).
Generally, the value of $\gamma \approx 2.3$, which is predicted by a model for a short-range impurity potential \cite{KramerPR-2005-43,HuckesteinRMP-1995-44}, has been observed in a number of works on VRH (see references in Ref.~\onlinecite{ArapovJLTP2016-25,GudinaPSSC2016-26}).

On the other hand, in the AlGaAs/GaAs heterostructure (a symmetrical modulation-doped GaAs quantum well bounded by Si $\delta$-doped AlGaAs layers on each side) the values of $\gamma$ are found to be $1.3 \pm 0.2$ with a perfect fit of longitudinal conductivity $\sigma_{xx}$ as a function of the scaling variable for the VRH regions at three filling factors  $\nu = 5$, 6 and 7 \cite{ZhaoPRB-2008-51}.

These values, which correspond to $\gamma = 4/3$ in the theories of classical percolation \cite{TrugmanPRB-1983-52,LeePRL-1993-75}, show that just classical percolation dominates scaling behavior in the samples due to the presence of the long range potential fluctuation caused by remote ionized impurities in AlGaAs. 

Cobaleda et al.\cite{CobaledaPRB-2014-80} measured the critical exponent $\gamma$ in bilayer graphene (encapsulated by $h$-BN) for a number of  PPTs, at different carrier densities tuned by the back gate voltage ($V_g$), both for negative and positive charge carriers.   
From the analysis of the longitudinal conductivity in the regime of variable range
hopping at different $V_g$, a set of estimates for $\gamma$ have been obtained with the mean value $\gamma = 1.25$ (0.96, 1.54). This value is entirely compatible with a classical percolation picture ($\gamma = 4/3$) and is definitely different from the value of $\gamma = 2.38$, which has been found in monolayer graphene \cite{BennaceurPRB-2012-48}.

In our system, the scaling in VRH regime is realized at a sufficiently large distance from the center of the Landau level: for $|\nu -\nu_c| \gtrsim 0.2$ (see Fig.~\ref{fig9:Xi-nu}), displaying  the divergence of localization length with the exponents $\gamma = 1.31 \pm 0.03$ at  $\nu_c \approx 1.5$  and $\gamma = 0.93 \pm  0.03$ at  $\nu_c \approx 2.5$. Thus, we deal with the hopping between localized states in the tails of Landau levels, that is within the scope of the laws of classical percolation outside the region of quantum tunneling for a long-ranged impurity potential of the remote In ions.

In our sample, two alternative scaling laws for PPT width $\nu_0(T)$ are valid for different regions of $\Delta \nu$, in analogy with Ref.~\onlinecite{TaoPRL-2007-50}: at $|\nu - \nu_c| \lesssim 0.1$  and within $0.2 \lesssim |\nu - \nu_c| \lesssim 0.35$ for VRH regime, that  is indicated in Fig.~\ref{fig9:Xi-nu} by the arrows, respectively.

\section{\label{sec:5}Conclusions}

We have measured the longitudinal and Hall resistivities in the quantum Hall regime at magnetic fields $B$ up to 9T and temperatures $T = (2.9-50)$K for a HgTe quantum well with inverted energy spectrum ($d_{QW} = 20.3$nm).
Temperature dependence of the plateau-plateau transitions width, $\nu_0(T)$, is studied and the actual scaling behavior $\nu_0(T) \sim T^{-\kappa}$ have been observed for the $1 \to 2$ plateau-plateau transition ($\nu_c = 1.5$) in a wide temperature range $T = 2.9-30$K. The extracted critical exponent $\kappa = 0.54 \pm 0.01$ is in quite good accordance with experimental data for other systems with a large-scale impurity potential.

A set of our experimental data on the temperature dependence of conductance in the minima associated with the Hall plateau regions may be successfully interpreted in terms of the variable-range hopping in the presence of a Coulomb gap. We have found that the hopping conductivity dominates in the regions of both first and second Hall plateaus, thus we used the theory of hopping of interacting electrons  to extract, in a straightforward way, the magnetic-field dependence of the localization length, $\xi(B)$. 
An analysis of the $\xi(B)$ dependence revealed that for the HgTe quantum well we deal with the hopping between localized states in the tails of Landau levels in the investigated range of fields and temperatures that corresponds to the region of classical percolation through a long-range impurity potential of the remote In ions.

The results we obtained suggest the possibility of implementing the scaling regime both for the QHE plateau-plateau transition ($|\nu - \nu_c| \lesssim 0.1$) and VRH regime within the Hall plateau regions ($0.2\lesssim |\nu - \nu_c|\lesssim  0.35$) in the 2D structures based on mercury telluride. 

Note that the temperature ranges where scaling laws are observed differs significantly for various materials: from liquid-helium and sub-liquid-helium temperatures for III-V structures to temperatures of ~100 K for single- and double-layer graphene. For the studied structure based on HgTe, the range extends to $T \approx 30$K due to large cyclotron and spin splittings of Landau levels because of the extraordinarily small effective mass and large $g$-factor.

\section{\label{sec:6}Appendix}

Here is a table of experimental results for critical exponent ($\kappa$) values extracted 
from the temperature dependences of the PPT width in  modulation-doped GaAs/AlGaAs heterostructures \cite{TaoPRL-2007-50,ZhaoPRB-2008-51,LiPRL-2005-61,KochPRL-1991-65,WeiPRB-1992-66,YooSSC-1994-67,KochSST-1995-68,HohlsPRL-2002-69,DodooJPCM-2014-70,Nakajima-2007-71,LiPRB-2010-72,XuebinPRB-2016-73}.

\begin{widetext}

\begin{table}[h]
\caption{\label{tab2:Kappa} The critical exponent ($\kappa$) values for modulation-doped GaAs/AlGaAs heterostructures. }
\begin{tabular}{|c|l|l|l|c|}
\hline
Structure & PPT & Value of $\kappa$ & Method & Ref.\\  
\hline
                        & $2 \to 1$ & 0.42 $T<0.2$K           &                                            & \\
GaAs/Al$_x$Ga$_{1-x}$As & $3 \to 2$ & $0.72\pm 0.2$ $T>0.75$K & $|\frac{d\rho_{xy}}{dB}|_{max}$ [0.02-5K]  &[\onlinecite{WeiPRB-1992-66}]\\
                        & $4 \to 3$ &                           &  &\\
\hline
                        & $3 \to 2$ & $0.68 \pm 0.04$           &     $\Delta B \sim  T^\kappa$  & \\
GaAs/Al$_x$Ga$_{1-x}$As & $4 \to 3$ & $0.72\pm 0.05$  & $|\frac{d\rho_{xy}}{dB}|_{max}$ [0.025-1K]  &[\onlinecite{KochPRL-1991-65}]\\
                        & $5 \to 4$ & $0.67 \pm 0.06$                        &  &\\
\hline
GaAs/Al$_{0.3}$Ga$_{0.7}$As  & $3 \to 2$, $4 \to 3$  & $0.5 \pm 0.03$ & $\Delta B \sim  T^\kappa$ [0.3-1.2K]&[\onlinecite{YooSSC-1994-67}]  \\
 & $5 \to 4$, $6 \to 5$ & $0.5\pm 0.03$  & $|\frac{d\rho_{xy}}{dB}|_{max}$   &\\
\hline
GaAs/Al$_x$Ga$_{1-x}$As & $4 \to 3$ & $0.62 \pm 0.04$ & $\Delta B \sim  T^\kappa$ [0.05 - 1K] &[\onlinecite{KochSST-1995-68}]\\
 & & $0.59 \pm 0.04$ & $\Delta B \sim  J^{\kappa/2}$ &\\
\hline
GaAs/Al$_x$Ga$_{1-x}$As & $2 \to 1$ & $0.66 \pm 0.02$ S1 & $\Delta \nu \sim  T^\kappa$ [0.05 - 1K] &[\onlinecite{HohlsPRL-2002-69}]\\
 & & $0.60 \pm 0.02$ S2 & & \\
 & & $0.62 \pm 0.02$ S3 & &\\
\hline
GaAs/Al$_{0.22}$Ga$_{0.78}$As & $2 \to 1$ & $0.64 \pm 0.09$ & $|\frac{d\rho_{xy}}{dB}|_{max}$ [0.3 - 1K] &[\onlinecite{HuangPE-2004}]\\
\hline
Al$_x$Ga$_{1-x}$As/Al$_{0.32}$Ga$_{0.68}$As & $6 \to 5$ & $0.58 - 0.49$ & $\Delta B \sim  T^\kappa$ [0.03 - 1K]&[\onlinecite{LiPRL-2005-61}]\\
$x<0.0085$ & $5 \to 4$& $0.58 - 0.50$ & $|\frac{d\rho_{xy}}{dB}|_{max}$  &\\
 &$4 \to 3$ & $0.57 - 0.49$ & &\\
\hline
GaAs/Al$_{0.35}$Ga$_{0.65}$As & $3 \to 2$ & $0.66 - 0.77$ & $\Delta B \sim  T^\kappa$ [1.7 - 4K] &[\onlinecite{TaoPRL-2007-50}]\\
 &$4 \to 3$ &  & $|\frac{d\rho_{xy}}{dB}|_{max}$ &\\
\hline
GaAs/Al$_{x}$Ga$_{1-x}$As & $6 \to 5$ & $0.72(0.74)$ & $\Delta \nu \sim  T^\kappa$ [0.05 - 1.2K] &[\onlinecite{ZhaoPRB-2008-51}]\\
 &$7 \to 6$ & 0.74(0.80) & $|\frac{d\rho_{xy}}{dB}|_{max}$ &\\
 &$8 \to 7$, $10 \to 8$ & $0.75\pm 0.05$ &  &\\
\hline
GaAs/Al$_{x}$Ga$_{1-x}$As & $1 \to 0$ & $0.79$ & $\Delta \nu \sim  T^\kappa$ [0.05 - 5K]   &[\onlinecite{Nakajima-2007-71}]\\
mesoscopic system &$3 \to 2$ & 0.54 &  &\\
\hline
Al$_x$Ga$_{1-x}$As/Al$_{0.32}$Ga$_{0.68}$As & $4 \to 3$ & $0.42$ $15 <T<120$mK & $\Delta B \sim  T^\kappa$ [0.03 - 1.2K]&[\onlinecite{LiPRB-2010-72}]\\
$x=0$ & $ $& $0.58$ $T>120$mK & $|\frac{d\rho_{xy}}{dB}|_{max}$  &\\
\hline
Al$_x$Ga$_{1-x}$As/Al$_{0.32}$Ga$_{0.68}$As & $4 \to 3$ & $0.42$ $10 <T<250$mK & $\Delta B \sim  T^\kappa$ [0.03 - 1.2K]&[\onlinecite{LiPRB-2010-72}]\\
$x=0.0021$ & $ $& $0.58$ $T>250$mK & $|\frac{d\rho_{xy}}{dB}|_{max}$  &\\
\hline
p-GaAs/Al$_x$Ga$_{1-x}$As & $3 \to 2$ & $0.52 \pm 0.01$ & $\Delta \nu \sim  T^\kappa$ [0.05 - 1K] &[\onlinecite{XuebinPRB-2016-73}]\\
 & $4 \to 3$ & $0.52 \pm 0.02$ & & \\
 & $5 \to 4$ & $0.53 \pm 0.02$ & &\\
\hline

\end{tabular}
\end{table}
\end{widetext}

In the Table~\ref{tab2:Kappa} the following abbreviations for a method of determination of the critical exponent from the experimental data on the Hall, $\rho_{xy}$, and the longitudinal, $\rho_{xx}$, resistivities  are used. The values of $\kappa$ have been found from the temperature dependences both of the slope of the steps between adjacent quantum Hall plateaus:
\begin{equation}
\left|\frac{d\rho_{xy}}{dB} \right|_{B=B_C} \equiv \left|\frac{d \rho_{xy}}{dB}\right|_{max} \sim T^{-\kappa},
\label{}
\end{equation} 
 and of the longitudinal resistance peak width at the PPT: 
\begin{equation}
\Delta B \sim T^\kappa .
\label{}
\end{equation} 

In Ref.~\onlinecite{KochSST-1995-68}  a scaling analysis of the current (J) dependence of the resistance peak width was also carried out:  
\begin{equation}
\Delta B \sim  J^{-\kappa/2} .
\label{}
\end{equation} 

It is seen from the Table that the discovered values of parameter $\kappa$ are in the main  concentrated at the range of (0.5 - 0.75).  Within the theoretical concepts for the large-scale impurity potential (see the text) it corresponds to a borderland between quantum tunnelling processes (\emph{genuine} scaling, $\kappa = 0.42$) and classical percolation regime ($\kappa = 0.75$).

Let's turn our attention on the results of Wei et al. \cite{WeiPRB-1992-66}   who have found that the $T$ dependence of $(d \rho_{xy}/dB)_{max}$ behaves like $T^{-0.42}$  in two low-mobility GaAs/Al$_x$Ga$_{1-x}$As heterostructures from the experiments down to $T =200$mK (see Table~\ref{tab2:Kappa}). It is similar to their earlier reported result for the In$_x$Ga$1-x$As/InP  heterostructure \cite{WeiPRL-1988-60} but at more lower temperatures. 

The 2DEG in the In$_x$Ga$_{1-x}$As/InP heterostructure is in the alloy In$_x$Ga$_{1-x}$As  layer and the potential fluctuations are therefore short ranged compared to the cyclotron radius (typically 100 \AA).   On the other hand, the 2DEG in the GaAs/Al$_x$Ga$_{1-x}$As heterostructures is in the GaAs layer, and the dominant scattering mechanism at low T is the remote ionized impurities away from the 2DEG layer. One should then expect smooth, long- range potential fluctuations \cite{EfrosSSC-1988-81,GoldAPL-1989-82}. The necessity to lower the temperature for detecting the "universal" scaling in GaAs/Al$_x$Ga$_{1-x}$As is  attributed just to the dominance of the long-range random potential.

Recently Li et al.\cite{LiPRL-2005-61} studied the dependence of the exponent $\kappa$  on $x$ for Al$_x$Ga$_{1 x}$As/Al$_{0.33}$Ga$_{0.67}$As heterostructures in a wide Al concentration range and have distinguished three regimes. 

For samples in the first regime ($x < 0.0065$), where the long-range potential for scattering on remote ionized impurities is the main one, $\kappa$ reaches $0.56-0.58$. For the second regime ($0.0065 < x < 0.016$), the probability of short-range alloy scattering becomes significantly higher, the transport has a quantum nature, and $\kappa = 0.42$ for all samples. Finally, at $x > 0.016$, $\kappa$ again increases to $0.57-0.59$ because of Al-atom clusterization resulting in a change in the character of disorder in the system (macroscopic inhomogeneities), thus breaking the universal scaling. 

It is assumed in Ref.~\onlinecite{LiPRL-2005-61} that quantum tunnelling processes (for the short-range impurity potential) are followed by classical processes (for the large-scale potential) with increasing disorder range. Due to the quantum-classical crossover effect the exponent $\kappa$ increases from 0.42 towards the classical value of 0.75. The fact that the $\kappa$ values obtained in the first and third regimes, which are  still  well below 0.75, show that the system is still away from an ideal classical percolation regime.

In their subsequent work \cite{LiPRB-2010-72}, extending temperature range from 1.2 K down to 1mK for Al$_x$Ga$_{1-x}$As/Al$_{0.32}$Ga$_{0.68}$As  heterostructures in a region of long-range disorder (for  x = 0 and 0.0021)  Li et al. have observed a crossover behavior from the high-temperature nonuniversal scaling regime to the low-temperature universal scaling regime with the temperature exponent $\kappa$ changing from $\kappa =0.58$ to 0.42, respectively (see Table ~\ref{tab2:Kappa}).

\end{document}